\documentclass[aps,prd,twocolumn,reprint,showpacs,floatfix]{revtex4-1}

\usepackage{graphics}
\usepackage{graphicx}
\usepackage{amssymb}
\usepackage{bm}
\usepackage{dcolumn}

\def\beq{ \begin{equation}}
\def\eeq{\end{equation} }
\def\bea{\begin{eqnarray}}
\def\eea{\end{eqnarray}}

\begin{document}

\title{Impact of internal bremsstrahlung on the detection of $\gamma$-rays from neutralinos}

\author{M.~Cannoni$^1$, M. E. G\'omez$^1$, M. A. S\'anchez-Conde$^{2,3}$, F. Prada$^4$ and O. Panella$^5$}
\affiliation{$^1$ Departamento de F\'isica Aplicada, Facultad de Ciencias Experimentales, Universidad de
Huelva, 21071 Huelva, Spain}
\affiliation{$^2$ Instituto de Astrof\'isica de Canarias (IAC), E-38200 La Laguna, Tenerife, Spain  }
\affiliation{$^3$ Departamento de Astrof\'isica, Universidad de La Laguna (ULL), E-38205 La Laguna, Tenerife, Spain  }
\affiliation{$^4$ Instituto de Astrof\'isica de Andaluc\'ia (CSIC), E-18008, Granada, Spain}
\affiliation{$^5$ Istituto Nazionale di Fisica Nucleare, Sezione di Perugia, Via Alessandro Pascoli,
06129, Perugia, Italy}

\begin{abstract}

We present a detailed study of the effect of internal bremsstrahlung photons in the context of the minimal supersymmetric standard models and their impact on $\gamma$-ray dark matter annihilation searches. We find that although this effect has to be included for the correct evaluation of fluxes of high energy photons from neutralino annihilation, its contribution is relevant only in models and at energies where the lines contribution is dominant over the secondary photons.
Therefore, we find that the most optimistic supersymmetric scenarios
for dark matter detection do not change significantly when including
the internal bremsstrahlung.
As an example, we review the $\gamma$-ray dark matter detection prospects of the Draco dwarf spheroidal galaxy for the MAGIC stereoscopic system
and the CTA project. Though the flux of high energy photons is enhanced by an order of magnitude in some regions of the parameter space, the expected fluxes are still much below the sensitivity of the instruments.

\end{abstract}



\maketitle


The minimal supersymmetric (SUSY) extension  of the standard model (MSSM)
provides a natural candidate for dark matter (DM) in the form of a neutral,
stable Majorana fermion, the lightest neutralino.
At present, large efforts are being carried out to detect
this SUSY DM by different methods, see~\cite{bertone} for reviews.

In the case of the current imaging atmospheric Cherenkov telescopes (IACTs),
the searches are based on the detectability of $\gamma$-rays coming from the
annihilation of the SUSY DM particles in the halo of galaxies~\cite{gammas}.
Neutralinos annihilate at the one loop level into photons through the processes~\cite{lines}
$\chi \chi \rightarrow \gamma \gamma$,
$\chi \chi \rightarrow Z \gamma$,
with almost monochromatic outgoing photons of energies
$E_\gamma\sim m_\chi$
$E_\gamma\sim m_\chi- {m_Z^2 /{4 m_\chi}}$,
respectively. Moreover, neutralino annihilation can produce a
continuum spectrum of secondary photons from hadronization and decay of
the annihilation products, mostly from neutral pion decay, which typically
dominates over the number monochromatic $\gamma$'s in
a large portion of the parameter space.
IACTs in operation like MAGIC, HESS, VERITAS~\cite{magic} or satellites-based
experiments like the Fermi satellite~\cite{fermi} play a very important role in this kind of DM searches.
For these experiments, dwarf spheroidal (dSph) galaxies around the Milky Way represent a good alternative target option
to e.g. the Galactic Center, already observed in $\gamma$-rays but with null DM detection so far~\cite{search}. 
Dsphs are DM dominated systems with inferred very high
mass-to-light ratios, and most of them are expected to be free from any other astrophysical source that might contribute 
to a possible $\gamma$-ray signal. Therefore, the detection of $\gamma$-rays from them would probably imply a successful 
DM annihilation detection.

Some of the present authors calculated  in Ref.~\cite{PRD1} the
expected $\gamma$-ray flux due to neutralino annihilation in the Draco
dSph for a typical IACT above 100 GeV. Draco is located at 80 kpc and
is one of the dwarfs with more observational constraints, which have
helped to better determine its DM density profile. 

The MAGIC telescope has already observed Draco in $\gamma$-rays in the context of DM searches~\cite{MAGICdraco}, 
but found no gamma signal above an energy threshold of 140 GeV. As a consequence, an
upper limit for the flux (2$\sigma$ level) was set to be 
$1.1 \times 10^{-11}$ph~cm$^{-2}$s$^{-1}$, assuming a power-law with spectral index $-1.5$ and a point-like source. 
This upper limit is 
${\mathcal O}(10^3 - 10^9 )$ above the values predicted by those SUSY models used in their analysis and therefore no constraints
could be put on the parameter space.  
Also the Fermi collaboration has recently reported their upper limits for a possible $\gamma$-ray annihilation signal from Draco at lower energies~\cite{fermidSphs}, given that no significant gamma emission was detected above 100 MeV.

Recently, in Ref.~\cite{IB1}, it was noted that the photons arising
from {\it internal bremsstrahlung} (IB) in some regions of the parameter space
can dominate the spectrum at energies near the neutralino mass.
IB~\cite{IB1,IB2} is commonly  referred to the emission of additional photons
in neutralino pair annihilation into charged final states, $\chi\chi \to X\bar{X}\gamma$
($X$ being a charged lepton, a quark, a $W$ boson or a charged Higgs) which is
an unavoidable electromagnetic radiative correction. In the Feynman diagrams
these photons can be attached to the external legs representing final state charged particles
or to the  propagator of the virtual charged particle exchanged by neutralinos:
the latter diagrams are at the origin of the hard photon spectrum of IB.

The aim of this brief report is
to quantitatively study the impact of the IB on the DM detection prospects for IACTs. As an example, we will revisit Draco updating the results obtained in Ref.~\cite{PRD1}, this time fully taking into account the IB contribution to the expected annihilation flux.  We note that a study of Draco including IB was already done in Ref.~\cite{DracoIB}, but only for a few benchmark points of the parameter space. Here, we will perform a wider exploration of the parameter space and will extract more general conclusions on the real importance of the IB for $\gamma$-ray DM searches.
%
%

The expected flux of photons with energy above the
threshold of the telescope, $F(E_\gamma >E_{\rm th})$, is given
by the product of the so-called astrophysical factor $J(\Psi)$ times the particle physics factor, namely $\Phi_{PP}(E_\gamma >E_{\rm th})$.
$J(\Psi)$ represents the integral
of the square of the dark matter density $\rho_{DM}$ along the
direction of observation $\Psi$ relative to the center of the DM halo, and depends on the PSF of the telescope. In the case of Draco, for instance, the authors in Ref.~\cite{PRD1} used a cusp and a core DM density profiles, built from the latest stellar kinematic observations together with a rigorous method of removal of interloper stars. In the same work, they also stressed the important role of the PSF, which is directly related to the angular resolution of the IACT and becomes crucial for a correct interpretation of a possible gamma signal due to neutralino annihilation. However, for the sake of simplicity, we will use here the value of $J(\Psi)$ integrated over the whole spatial extent of the source as the value of the astrophysical factor. This value, that does not depend on the PSF any longer, can be well approximated by
$\overline{J}= \frac{1}{4\pi D^2}\int_{V} \rho_{DM}^2(r)~dV$,
with $D$ the distance from the Earth to the center of the DM halo and
$r$ the galactocentric distance inside it. For Draco we take a value of $\overline{J} = 3.7\times 10^{17}$ Gev$^2$ cm$^{-5}$, which was calculated using the cuspy DM density profile given in Ref.~\cite{PRD1}.  
%
%
%
%
\begin{table*}[htbp!]
\caption{mSUGRA models used in Fig.~\ref{fig:IB}.
The values of $m_{0}$, $m_{1/2}$, $A_{0}$, $m_{\tilde{\chi}}$ are in GeV,
the sign of $\mu$ is positive. The units of $\langle \sigma_{\chi\chi} v \rangle$ are
cm$^{3}$ s$^{-1}$, those of the $f$'s, defined in Eq.~(\ref{fsusy}),  are GeV$^{-2}$ cm$^{3}$ s$^{-1}$. }
\begin{ruledtabular}
\begin{tabular}{ c  c  c  c  c  c  c  c c c c c}
 & $\tan\beta$ & $m_{0}$  & $m_{1/2}$ & $A_{0}$ & $m_{\tilde{\chi}}$ &
$\Omega_\chi h^2$ & $\langle \sigma_{\chi\chi} v \rangle \cdot 10^{29}$ & $f_{sec}\cdot 10^{32}$ & $f_{lines}\cdot 10^{32}$ & $f_{IB}\cdot 10^{32}$ & $f_{SUSY}\cdot 10^{32}$  \\
\hline
A & $18$ & $127$ & $459$ & $-135$ & 187.6 & 0.092 & 29  &0.008 & 0.018 & 0.079 & 0.1\\
B & $52$ & $982$ & $1377$ & $725$ & 597.6 & 0.092 & 2600 & 0.72 &10$^{-5}$ &10$^{-5}$ & 0.72 \\
C & $17$ & $2200$ & $430$ & $805$ & 162.8 & 0.098 & 2225 &0.04 &0.06 &0.02 & 0.12 \\
D & $51$ & $8940$ & $2218$ & $-4221$ & 918.2 & 0.099  &1203  &0.3 &0.003 &0.017  & 0.32  \\
E & 5    & 110       & 530   &  $-600$   & 218.4 &  0.1  & 11.2  & 0.0014 & 0.014     &0.073       &0.088
\end{tabular}
\end{ruledtabular}
\label{tab}
\end{table*}

As for $\Phi_{PP}(E_\gamma >E_{\rm th})$, which in the following we call $f_{SUSY}$,
it includes all the particle physics informations and is made up by
the contribution of the continuum spectrum (secondaries and IB photons)
and the monochromatic photons (lines):
\begin{eqnarray}
&f_{susy}
=f_{cont}+f_{lines},&\cr
&f_{cont}
=\left( \sum_f B_f \int_{E_{th}}^{m_{\chi}} \frac{dN^{f}_{\gamma}}{dE_{\gamma}}dE_{\gamma} \right)
\frac{\langle \sigma_{\chi\chi} \,v\rangle} {2 m_{\chi}^2} = f_{sec} + f_{IB},&\cr
&f_{lines} = 2 \frac{\langle \sigma_{\gamma\gamma} \,v \rangle}     {2 m_{\chi}^2} +
\frac{\langle \sigma_{Z\gamma} \,v\rangle           }{2 m_{\chi}^2}.&
\label{fsusy}
\end{eqnarray}
Here ${dN^{f}_{\gamma}}/{dE_{\gamma}}$ is the differential yield of photons per annihilation to
the final state $f$ with branching ratio $B_f$. The factor in parenthesis is
thus $n_{\gamma} (E_{\gamma} > E_{th})$, the total number of photons per annihilation
with energy greater than the threshold energy, $\langle \sigma_{\chi\chi} \,v\rangle$
is the thermal averaged total neutralino annihilation cross,  
$\langle \sigma_{\gamma\gamma} \,v \rangle$ and 
$\langle \sigma_{Z\gamma} \,v\rangle$ the cross sections for annihilation into lines 
and $m_{\chi}$ the neutralino mass.
\begin{figure*}[htbp!]
\includegraphics*[scale=0.6]{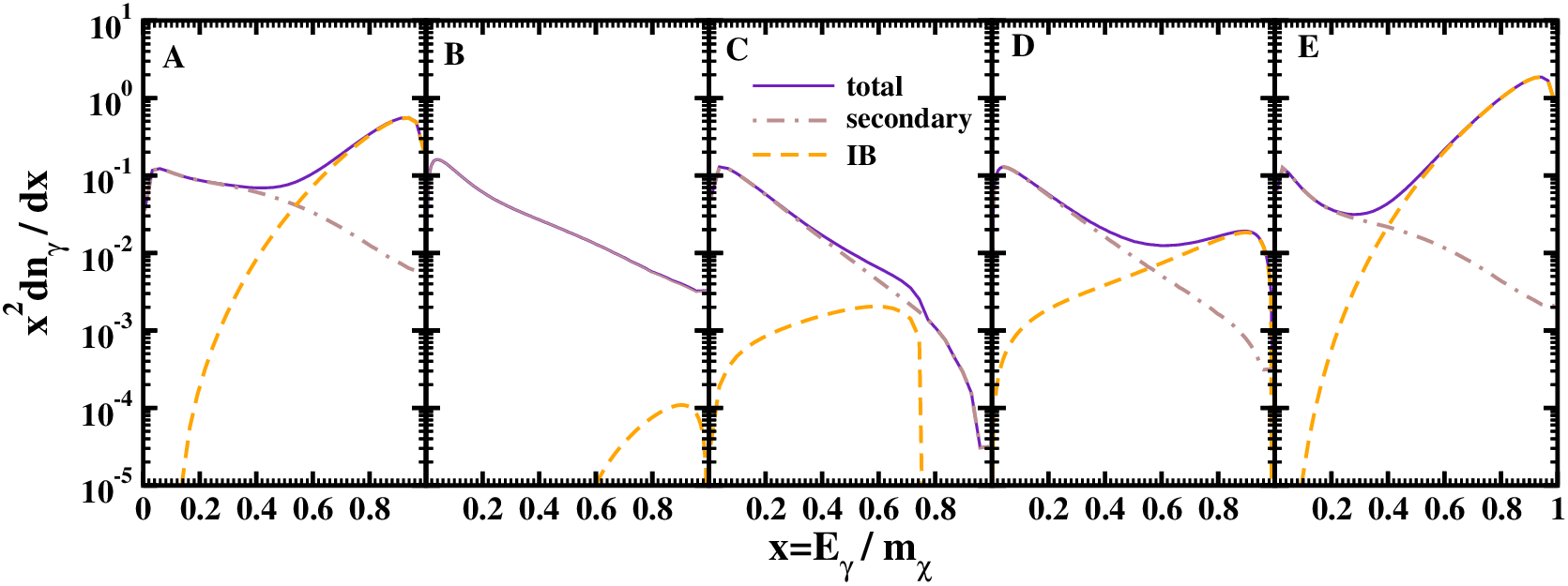}
\caption{Continuum $\gamma$-rays spectra for the models in Table~\ref{tab}.
The solid lines indicate the total yield, the dot-dashed lines the contribution of secondary photons,
while dashed lines correspond just to the IB contribution. }
\label{fig:IB}
\end{figure*}

As in Ref.~\cite{PRD1} we assume that the neutralino is the main component of the DM present
in the Universe with abundance inside the cosmologically favored interval
$0.09 <\Omega_\chi h^2 < 0.13$ (the most recent
WMAP~\cite{WMAP} interval at $3\sigma$ is $0.094 <\Omega_{DM} h^2 < 0.128$).
We further require that SUSY models satisfy the LEP bounds on Higgs and chargino masses,
$m_h >114$ GeV and $m_{\chi^+} > 103.5$ GeV, and constraints from $b\to s \gamma$
as explained in Refs.~\cite{PRD1,mario}.
We first consider minimal supergravity (mSUGRA) models, where the soft terms of the MSSM
are taken to be universal at the gauge unification scale $M_{GUT}$.
The effective theory at energies
below $M_{GUT}$ is thus determined by four universal parameters: 
the common scalar mass
$m_0$, 
the gauginos mass $m_{1/2}$,  
the trilinear couplings
$A_0$, and the ratio of the Higgs vacuum expectation values,
$\tan\beta$. In addition, the
minimization of the Higgs potential leaves undetermined the sign of
the Higgs mass parameter $\mu$ that we take positive.
For the numerical computation of IB effects we use {\textsl{DarkSusy 5.0.5}}~\cite{Gondolo:2004sc}
which relies on {\textsl{ISASUGRA 7.78}} for the renormalization
group equation evolution of parameters.

The amount of photons coming from IB depends on the
annihilation channels and therefore it depends on the nature of the neutralino
in every particular SUSY model. We thus display
in Fig.~\ref{fig:IB} the differential yield of photons
for some representative points of the mSUGRA parameter space specified in Table~\ref{tab}
where the distinct contributions to $f_{SUSY}$ integrating the number of photons above
$E_{th} =100$ GeV can be read.
Model A is in the {\it stau coannihilation region} of the mSUGRA parameter space: the mass
of the lightest stau is $m_{\tilde{\tau}} =195$ GeV very close to $m_{\chi} =188$ GeV.
Neutralino pair annihilation in $\tau^+ \tau^- $ mediated by $t$-channel exchange of stau has the highest
annihilation cross section thus IB is relevant. Here $f_{IB}$ is the dominant contribution being
10 and 4.4 times greater than $f_{sec}$ and $f_{lines}$.
Model B is in the {\it funnel or resonances region}: the mass of the CP-odd neutral Higgs is
$m_A = 1211$ GeV while $m_{\chi} =598$ GeV, thus $m_A \simeq 2 m_\chi^{0}$
and WMAP bounds are satisfied due to neutralino pair annihilation
into fermion through $s$-channel exchange of heavy neutral Higgs bosons.
In this case no photon line can be attached to the virtual particles
and the IB yield is negligible.
Model C is in the {\it focus point or hyperbolic branch region}. The mass
of the lightest chargino is $m_{\chi^\pm} = 212$ GeV, not much bigger than $m_{\chi} = 163$ GeV
and neutralino pairs annihilate into $W^+ W^-$ through
$t$-channel chargino exchange. The IB yield is small
because $m_{\chi}$ is not much greater than $m_W$ and photons energy has a cut off
which corresponds to the kinematic endpoint
$x = 1-{{m_W}^2}/{m_{\chi}^2}\sim 0.75$. Here $f_{lines}$ is bigger
than $f_{sec}$ and $f_{IB}$.
Model D is another example in the focus point region. The mass of the lightest
chargino is $m_{\chi^\pm} = 954$ GeV, almost
degenerate with $m_{\chi} = 918$ GeV.
Neutralino pairs annihilate into $W^+ W^-$ through
$t$-channel chargino exchange as in C but in this case $m_{\chi} \gg m_W$ thus IB photons
contribution is more important and have endpoint at the neutralino mass: here $f_{SUSY}$
is dominated by $f_{sec}$.
The point E is similar to point A but with even more pronounced IB contribution
(here $m_h =113$ GeV). $f_{IB}$ is a factor 50 and 5 greater than $f_{sec}$ and $f_{lines}$ respectively
but the the total $f_{SUSY}$ is the smaller one.
\begin{figure*}[t!]
\includegraphics*[scale=0.6]{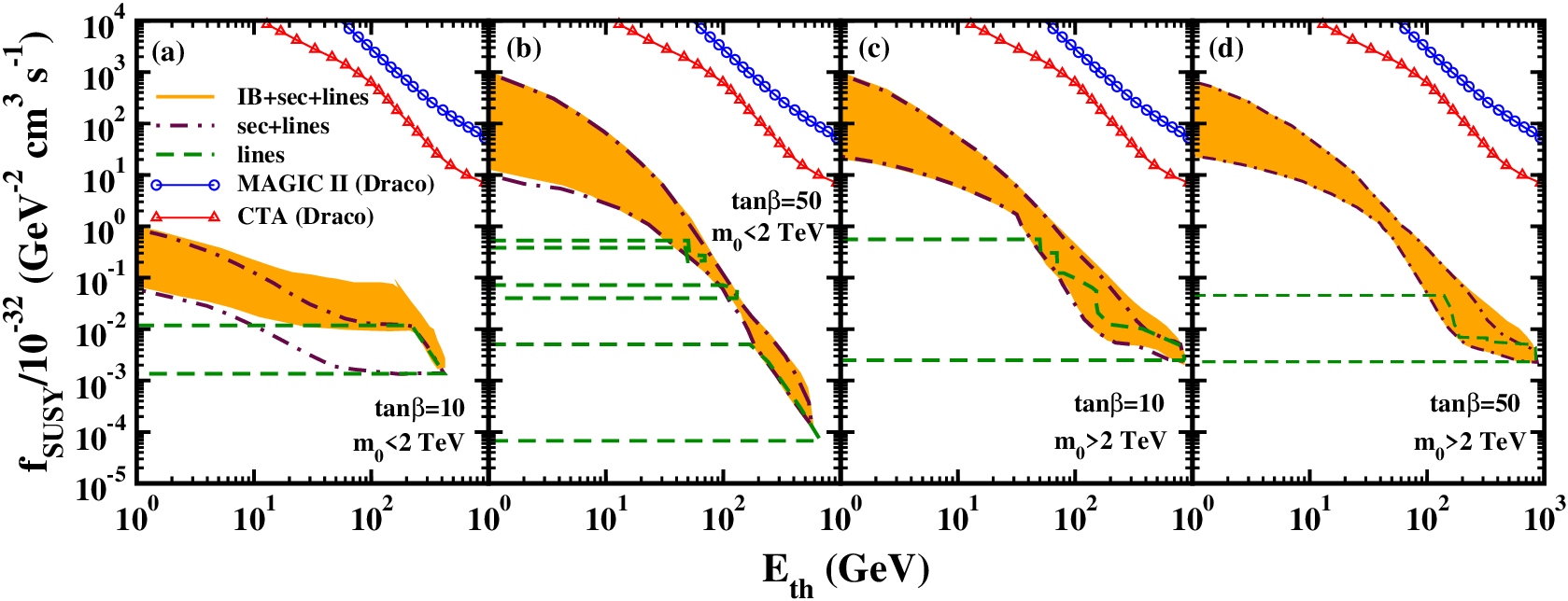}
\caption{The particle physics factor $f_{SUSY}$, Eq.~(\ref{fsusy}), versus $E_{th}$,
energy threshold of the detector.
We set $A_0 =0$ and $\mu$ positive while  $m_0$ and $m_{1/2}$ have values such that the
mSUGRA point predicts the neutralino relic density  inside the WMAP bounds
satisfying all the phenomenological constraints. Also plotted are the predicted sensitivity lines of MAGIC II and CTA 
for Draco corresponding to 50 hours of observation time and a 5$\sigma$ detection level. }
\label{fig:fsusy}
\end{figure*}

\begin{figure*}[htp!]
\includegraphics*[scale=0.6]{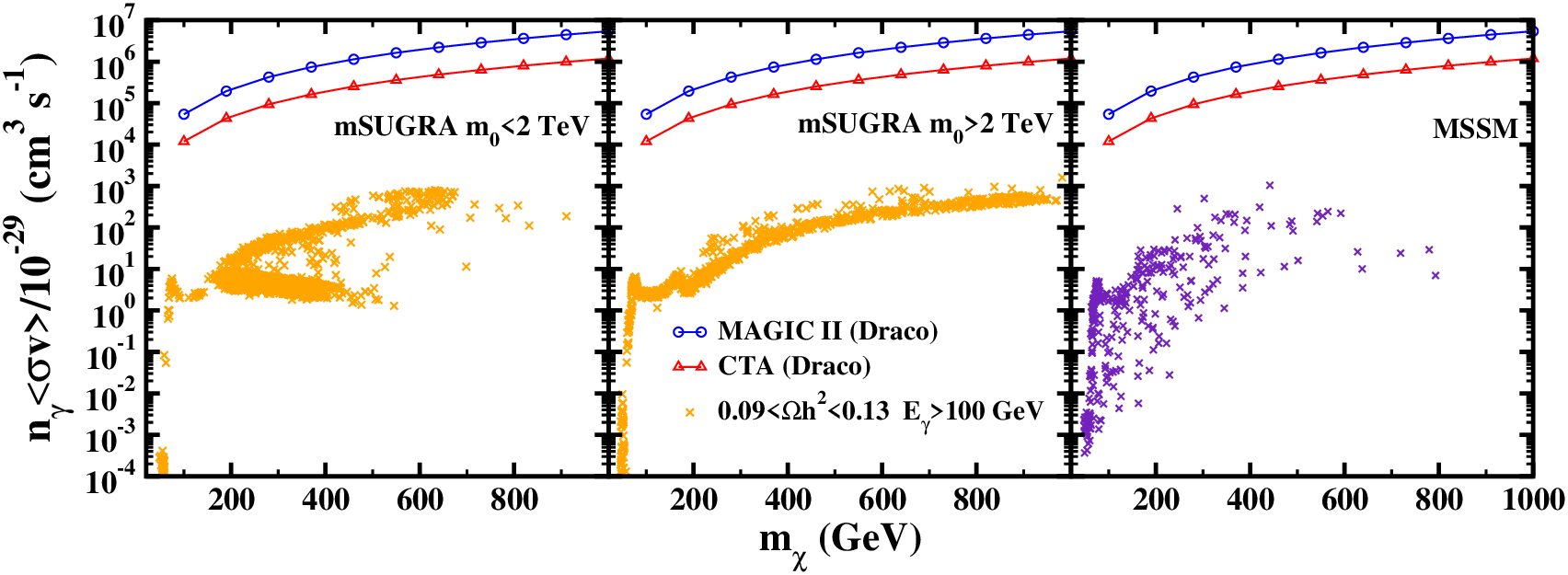}
\caption{Scatter plot of $n_\gamma (E_{\gamma} > 100$ GeV)$\times\left\langle \sigma_{\chi \chi}v\right\rangle$
versus $m_{\chi}$ for continuum photons in mSUGRA and general MSSM. Also plotted are the predicted sensitivity lines of MAGIC II and CTA 
for Draco assuming $E_{th} =100$ GeV, 50 hours of observation time and a 5$\sigma$ detection level.}
\label{fig:scat}
\end{figure*}
These ratios are quite general as
can be seen in Fig.~\ref{fig:fsusy} where  we present a larger scan on the parameter space
of $f_{SUSY}$ versus the threshold energy setting $A_0=0$ and two values of
$\tan\beta$, 10 and 50. In panel $(a)$  where points are
in the stau coannihilation region we can appreciate
the largest contribution of IB, as shown by the points A and E.
The absence of IB photons of the point B is evidenced by
the panel $(b)$ where points are in the funnel region,
while the panels $(c)$ and $(d)$ have points mostly in the hyperbolic branch
and share properties with the points C and D.
We also plot in those panels the sensitivity lines for Draco of the MAGIC telescopes in stereoscopic mode~\cite{magicII} 
and of the CTA project~\cite{cta} as given by Montecarlo simulations for 50 hours of observation time and a 5$\sigma$ detection level. 
We see that even including IB for a typical value of the threshold $E_{th} =100$ GeV, the theoretical predictions are at least three orders of magnitude below the detection limit.
%

We further perform a scan
on more general parameter space both in mSUGRA and in a general MSSM with random  soft terms
at the electroweak scale 
(see Ref.~\cite{PRD1} for details on the scanned parameter space).
In Fig.~\ref{fig:scat} we plot
the quantity $n_{\gamma}(E_{\gamma} >E_{th}) \times \langle \sigma_{\chi\chi} v\rangle$ 
versus $m_\chi$ together with the MAGIC II and CTA sensitivity lines assuming $E_{th} =100$ GeV: as above
we see that a boost of at least three and two orders of magnitude respectively
is needed to reach the detection line. We note, however, that the effect of substructures in the dwarf may enhance the annihilation flux importantly (see e.g. Ref.~\cite{pieri}).


In summary, we have reported a detailed study of the internal bremsstrahlung contribution
to the expected $\gamma$-ray flux from neutralino annihilation. We have found that although this effect has to be included for evaluation of fluxes of high energy photons from neutralino annihilation, its contribution is relevant only in models and at energies where the lines contribution
is dominant over the secondary photons. As a result, the most optimistic particle physics scenarios for DM detection (which typically correspond to those where most of the flux is given by secondary photons) will not change substantially. On the other hand, being typically the IB yield at most an order of magnitude greater than the lines yield, the net increase on absolute flux is of the same order. As an example of the impact of the IB on previous works on DM search, we recalculated the DM detection prospects of the Draco dwarf galaxy for the MAGIC telescopes, updating a previous work by some of the authors. We find that though the effect can rise the flux by e.g. an order of magnitude at 100 GeV, the predicted fluxes are still at least three orders of magnitude below the sensitivity of the instrument both in mSUGRA SUSY scenario and in the general MSSM. The same is applicable to other IACTs, given their roughly similar sensitivities at those energies.


The authors acknowledges support from the project
P07FQM02962 funded by "Junta de Andalucia",
the Spanish MICINN-INFN(PG21) projects FPA2009-10773, FPA2008-04063-E
and MULTIDARK project of Spanish MICINN
Consolider-Ingenio 2010: CSD2009-00064.


\begin{thebibliography}{99}


\bibitem{bertone}
G.~Bertone, D.~Hooper and J.~Silk,
Phys.\ Rept.\  {\bf 405}, 279 (2005);
C.~Munoz,
Int.\ J.\ Mod.\ Phys.\  A {\bf 19}, 3093 (2004).

\bibitem{gammas}
  Y.~B.~Zeldovich, A.~A.~Klypin, M.~Y.~Khlopov and V.~M.~Chechetkin,
  Sov.\ J.\ Nucl.\ Phys.\  {\bf 31}, 664 (1980)
  [Yad.\ Fiz.\  {\bf 31}, 1286 (1980)];
  J.~Silk and M.~Srednicki,
  Phys.\ Rev.\ Lett.\  {\bf 53}, 624 (1984).



\bibitem{lines}
L.~Bergstrom and P.~Ullio,
Nucl.\ Phys.\  B {\bf 504}, 27 (1997);
Z.~Bern, P.~Gondolo and M.~Perelstein,
Phys.\ Lett.\  B {\bf 411}, 86 (1997);
P.~Ullio and L.~Bergstrom,
Phys.\ Rev.\  D {\bf 57}, 1962 (1998).


\bibitem{magic}
  E.~Lorenz  [MAGIC Collaboration],
  New Astron.\ Rev.\  {\bf 48}, 339 (2004);
%
  J.~A.~Hinton  [The HESS Collaboration],
  New Astron.\ Rev.\  {\bf 48}, 331 (2004);
%
  T.~C.~Weekes {\it et al.},
  Astropart.\ Phys.\  {\bf 17}, 221 (2002).



\bibitem{fermi}
  N.~Gehrels and P.~Michelson,
  Astropart.\ Phys.\  {\bf 11}, 277 (1999).

\bibitem{search}
  F.~Aharonian {\it et al.}  [The HESS Collaboration],
  Astron.\ Astrophys.\  {\bf 425}, L13 (2004);
  J.~Albert {\it et al.}  [MAGIC Collaboration],
  Astrophys.\ J.\  {\bf 638}, L101 (2006);
  F.~Aharonian {\it et al.}  [H.E.S.S. Collaboration],
  Phys.\ Rev.\ Lett.\  {\bf 97}, 221102 (2006)
  [Erratum-ibid.\  {\bf 97}, 249901 (2006)].

\bibitem{PRD1}
M.~A.~Sanchez-Conde, F.~Prada, E.~L.~Lokas, M.~E.~Gomez, R.~Wojtak and M.~Moles,
Phys.\ Rev.\  D {\bf 76}, 123509 (2007).


\bibitem{MAGICdraco}
  J.~Albert {\it et al.}  [MAGIC Collaboration],
  Astrophys.\ J.\  {\bf 679}, 428 (2008).

\bibitem{fermidSphs}
  A.~A.~Abdo {\it et al.},
  Astrophys.\ J.\  {\bf 712}, 147 (2010).

\bibitem{IB1}
T.~Bringmann, L.~Bergstrom and J.~Edsjo,
JHEP {\bf 0801}, 049 (2008).

\bibitem{IB2}
L.~Bergstrom,
Phys.\ Lett.\  B {\bf 225}, 372 (1989);
R.~Flores, K.~A.~Olive and S.~Rudaz,
Phys.\ Lett.\  B {\bf 232}, 377 (1989);
L.~Bergstrom, T.~Bringmann, M.~Eriksson and M.~Gustafsson,
Phys.\ Rev.\ Lett.\  {\bf 95}, 241301 (2005);
V.~Barger, Y.~Gao, W.~Y.~Keung and D.~Marfatia,
Phys.\ Rev.\  D {\bf 80}, 063537 (2009).

\bibitem{DracoIB}
T.~Bringmann, M.~Doro and M.~Fornasa,
JCAP {\bf 0901}, 016 (2009).

\bibitem{WMAP}
  D.~Larson {\it et al.},
  arXiv:1001.4635 [astro-ph.CO].

\bibitem{Gondolo:2004sc}
P.~Gondolo, J.~Edsjo, P.~Ullio, L.~Bergstrom, M.~Schelke and E.~A.~Baltz,
JCAP {\bf 0407}, 008 (2004);
  F.~E.~Paige, S.~D.~Protopopescu, H.~Baer and X.~Tata,
  arXiv:hep-ph/0312045.

\bibitem{mario}
  M.~E.~Gomez, T.~Ibrahim, P.~Nath and S.~Skadhauge,
  Phys.\ Rev.\  D {\bf 72}, 095008 (2005);
{\it ibid.} {\bf 74}, 015015 (2006);
 M.~E.~Gomez, G.~Lazarides and C.~Pallis,
  Nucl.\ Phys.\  B {\bf 638}, 165 (2002),
  Phys.\ Rev.\  D {\bf 67}, 097701 (2003);
  M.~Cannoni and O.~Panella,
  Phys.\ Rev.\  D {\bf 81}, 036009 (2010).

\bibitem{magicII}
  P.~Colin {\it et al.} [MAGIC collaboration],
  arXiv:0907.0960. 

\bibitem{cta}
  M.~Doro [CTA collaboration],
  arXiv:0908.1410. 

\bibitem{pieri}
  L.~Pieri, A.~Pizzella, E.~M.~Corsini, E.~D.~Bonta' and F.~Bertola,
  Astron.\ Astrophys.\  {\bf 496}, 351 (2009).




\end{thebibliography}
\end{document}